\documentclass[twocolumn]{aastex62}

\graphicspath{{./}}

\received{\today}
\revised{\today}
\accepted{\today}

\submitjournal{ApJ}

\shorttitle{identification of Kelvin-Helmholtz vortices}
\shortauthors{Settino et al.}

\begin{document}

\title{Kinetic features for the identification of Kelvin-Helmholtz vortices in \textit{in situ} observations}

\correspondingauthor{Adriana Settino}
\email{adriana.settino@unical.it}

\author{A. Settino}
\affiliation{Dipartimento di Fisica, Universit\`a della Calabria, 87036 Rende (CS), Italy}
\affiliation{Swedish Institute of Space Physics, Box 537 SE--751 21 Uppsala, Uppsala, Sweden}

\author{D. Perrone}
\affiliation{ASI -- Italian Space Agency, via del Politecnico snc, 00133 Rome, Italy}

\author{Yu. V. Khotyaintsev}
\affiliation{Swedish Institute of Space Physics, Box 537 SE--751 21 Uppsala, Uppsala, Sweden}

\author{D. B. Graham}
\affiliation{Swedish Institute of Space Physics, Box 537 SE--751 21 Uppsala, Uppsala, Sweden}

\author{F. Valentini} 
\affiliation{Dipartimento di Fisica, Universit\`a della Calabria, 87036 Rende (CS), Italy}

\begin{abstract}

The boundaries identification of Kelvin-Helmholtz vortices in observational data has been addressed by searching for single-spacecraft small-scale signatures.
A recent hybrid Vlasov-Maxwell simulation of Kelvin-Helmholtz instability  has pointed out clear kinetic features which uniquely characterize the vortex during both the nonlinear and turbulent stage of the instability. 
We compare the simulation results with \textit{in situ} observations of Kelvin-Helmholtz vortices by the Magnetospheric MultiScale satellites.
We find good agreement between simulation and observations. In particular, the edges of the vortex are associated with strong current sheets, while the center is characterized by a low value for the magnitude of the total current density and strong deviation of the ion distribution function from a Maxwellian distribution.
We also find a significant temperature anisotropy parallel to the magnetic field inside the vortex region and strong agyrotropies near the edges.
We suggest that these kinetic features can be useful for the identification of Kelvin-Helmholtz vortices in \textit{in situ} data. 
\end{abstract}


\section{Introduction} \label{sec:intro}
Coherent structures, such as flux tubes and vortices, are some of the main features in both fluids and plasmas, whose presence is strongly associated with the development of turbulence. Turbulent phenomena play an important role in plasma transport, energy transfer across different scales and dissipation mechanisms beyond the inertial range where they can eventually take place. Studies on both solar-wind and near-Earth plasma have shown that ion scales are characterized by strong magnetic discontinuities \citep{retino2007,greco2016,perrone2016,perrone2017,perrone2020,wang2019}, which are connected through spatial scales from ion to electron scales.
Moreover, vortices have also been identified at larger scales, mainly associated to the Kelvin-Helmholtz (KH) instability, even if the identification of such vortices in \textit{in situ} observations is not straightforward.

The KH instability is an ubiquitous phenomenon that can develop, in both ordinary fluids and plasmas, when a shear flow exists. 
For instance, an unstable configuration is found when the velocity jump $\Delta u$ is larger than the component of the Alfv\'en velocity parallel to the bulk flow \citep{chandrasekhar1961}.
Shear flows have been observed, for instance, at the interaction region between fast and slow solar wind \citep{bruno2013} and both teories and simulations have shown that such sites are good candidates for the KH instability to grow \citep{korzhov1985,ismayilli2018}. In addition, in such regions, the wave-particle interaction with the non-uniform velocity field can produce small-scale fluctuations leading to the dissipation of the waves \citep{pezzi2017,valentini2012,valentini2017}.
KH instability has been observed in different plasma environments, such as in the solar corona at the edge of coronal mass ejections \citep{foullon2011} and at the planetary magnetospheres \citep{kivelson1995,fairfield2003, hasegawa2004a, hasegawa2006,foullon2010}. In particular, at the Earth's magnetopause the interaction with the shocked solar wind frequently leads to the formation of a so-called low latitude boundary layer and to the generation of KH vortices which propagate along the flanks of the magnetosphere and further towards the tail.

The KH instability, during its nonlinear and turbulent phases, can lead to the formation of thin current sheets which are possible sites for magnetic reconnection. Such coupling between KH instability and magnetic reconnection has been investigated by means of numerical simulations in both magnetohydrodynamics (MHD) and kinetic frameworks \citep[see i.e.][]{nakamura2013,faganello2017,franci2020}, as well as in observational data. Indeed, the Magnetospheric MultiScale (MMS) satellites are providing a deeper knowledge of the kinetic dynamics for KH instability \citep[see i.e.][]{eriksson2016,li2016,sorriso2019}. 
Nonetheless, several KH events has been observed during previous satellites missions, namely Cluster and THEMIS, at both flanks of the magnetosphere \citep[see][and references there in]{hwang2012}, providing also a statistical analysis of the dawn-dusk asymmetry \citep{henry2017,kavosi2015}.

Since KH instability plays a central role in several phenomena in space plasmas and, especially, in the context of near-Earth environment, it is crucial to identify KH vortices in order to better understand small-scale plasma dynamics. However, if the detailed study of KH vortices in numerical simulations is straightforward, due to the knowledge of both temporal evolution and spatial behavior, the identification of KH vortices in real space data, as collected by spacecraft, is very hard since only one point in space-time is provided and no information about the trajectory inside the vortex are available \textit{a priori}.

The main guidelines for the identification of KH vortices in observational data have been provided by \cite{hasegawa2004a}. Besides the observation of quasi-periodic fluctuations, a rotating pattern can be displayed by the hodograms of the velocity and/or magnetic perturbations. However, this vortical motion of the flux tubes can be clearly captured when the distances among the four satellites is large, as in the case of Cluster, whose average distance, during the detection of KH instability, was about $2000$~km. In such a case, it is impossible to study turbulent dynamics and all the related phenomena which characterize kinetic scales.
Moreover, MHD simulations of KH instability in the Earth's magnetospheric--like environment suggested a specific pattern to use for the identification of highly rolled-up vortices in \textit{in situ} measurement: the presence of lower density and faster than magnetosheath plasma regions \citep{takagi2006}. Nonetheless, this feature does not uniquely identify KH rolled-up vortices, but can also  be a signature of different phenomena as for example magnetosheath jets \citep{plaschke2014}.

Among the main quantities used for the identification of KH vortices, there are also the vorticity vector and the pressure minimum which are, however, inevitably affected by the choice of a threshold \citep[][]{hussain1987, hunt1988}.
In order to overcome such issues, mathematical techniques have also been developed for the identification of vortices in both fluids \citep[see][for a review]{jeong1995,kida1998} and magnetized plasmas \citep{cai2018}. 
Nonetheless, it is worth noting that these criteria have also some important limitions: i) the necessity of a multi-spacecraft analysis because they are based on the estimation of eigenvalues for the gradient of velocity (or magnetic) field vector; ii) the strong dependence of the vortex dimension on the relative distance between the spacecrafts.

The aim of our work is to provide new quantities that can be used as guidelines for the identification of the boundaries of KH vortices using only single-spacecraft measurements.
In this paper, we present a comparison of small-scale signatures observed in both simulation and \textit{in situ} data for a KH event. 
Recent hybrid Vlasov-Maxwell simulations have shown the presence of clear kinetic features in the vortices during both the nonlinear and turbulent stage of KH instability \citep{settino2020}.
Here, we investigate the presence of those features in a KH event observed by MMS during a period of northward interplanetary magnetic field. Leading and trailing edges of KH vortices have already been identified for this event by \cite{hwang2020} and \cite{kieokaew2020}. 
The paper is organized as follows: in Section \ref{sec:HVM} we discuss hybrid Vlasov-Maxwell simulation of KH instability and we present the features that identify the KH vortices during the nonlinear regime; in Section \ref{sec:MMS} we describe these same quantities as observed in the MMS data; in Section \ref{sec:discussion} we discuss and compare the results from both simulation and observational data, pointing out kinetic signatures in the distribution functions and, finally, in Section \ref{sec:summary} we provide conclusions.

\section{KH vortices in Vlasov simulations} \label{sec:HVM}

\begin{figure*}[ht]
    \centering
    \includegraphics[width=\textwidth]{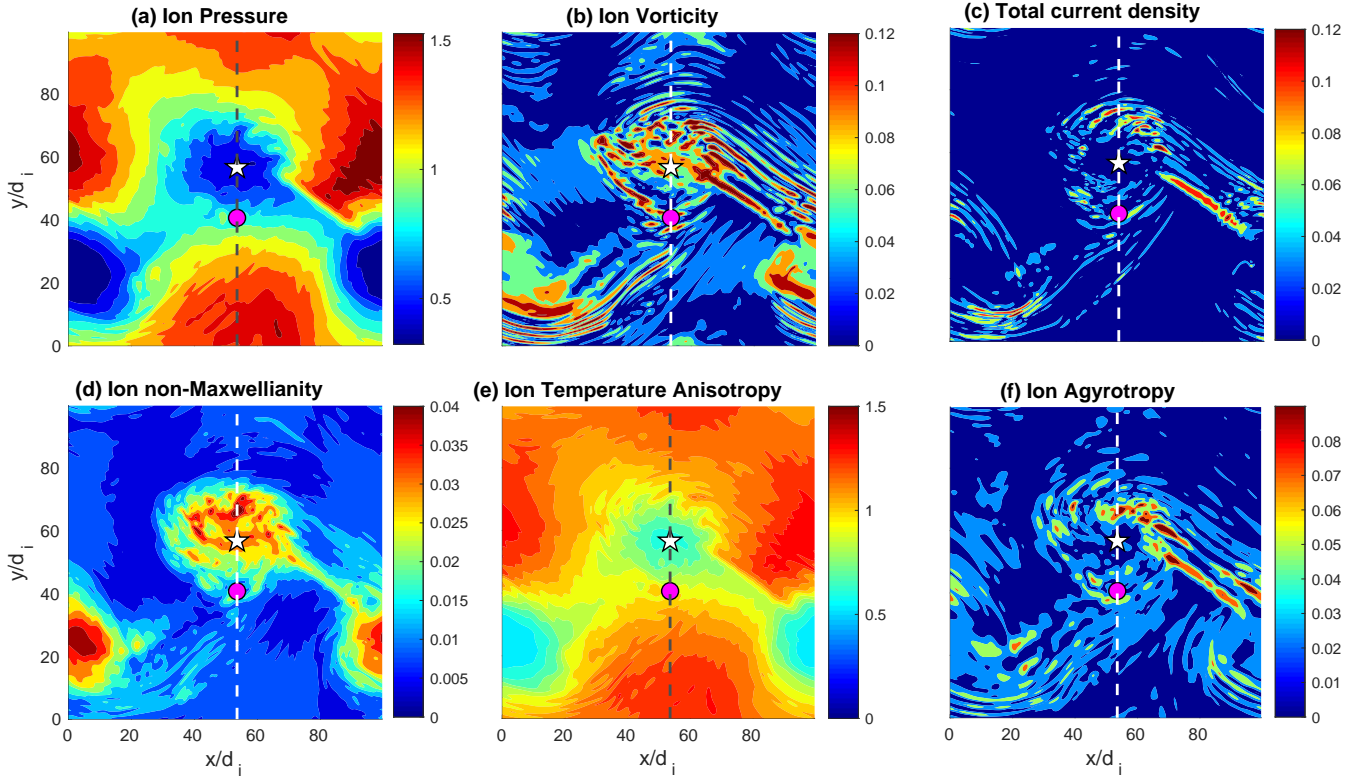}
    \caption{Two-dimensional contour plots for the HVM simulation at a fixed time during the turbulent stage of the KH instability. (a) Ion kinetic pressure, $P_i=n_i T_i$ (in normalized code units); (b) magnitude of the ion vorticity, $|\omega_i|$, and (c) total current density, $|{\bf j}|$; (d) ion non-Maxwellianity, $\epsilon_M$; (e) ion temperature anisotropy, $T_\perp/T_\parallel$; and (f) ion agyrotropy, $\sqrt{Q}$. The vertical dashed lines indicate a one-dimensional path in the two-dimensional box domain which crosses the vortex in the center. The spatial behavior of these quantities, along the $1$D cut, will be shown in Figure~\ref{fig:cut}. The white star and magenta circle in each iso-contour indicate the two spatial positions selected for investigating the ion distribution function (see Figure~\ref{fig:vdf}).} 
    \label{fig:contour}
\end{figure*}

When KH instability develops at ion scales, meaning that the thickness of the shear layer is of the order of the characteristic ion scales, such as at the Earth's magnetopause, kinetic effects come into play and a kinetic numerical approach can better describe the dynamics of the instability compared with a fluid approach \citep{nakamura2013,henri2013,karimabadi2013,rossi2015}. An important step for a correct description of the KH instability is to set up the equilibrium unperturbed state. However, the choice is not trivial since this setting has strong consequences for the onset of the instability. The simplest choice is to use a shifted Maxwellian distribution function, with the associated moments varying in space \citep{umeda2014}. However, a shifted Maxwellian is not an equilibrium distribution in a kinetic framework and can give rise to spurious oscillations of the order of the ion gyroperiod, which can affect the dynamics of the instability itself.

An exact stationary solution for the hybrid Vlasov-Maxwell system of equations, which describes a magnetized shear flow, has recently been found by \cite{Malara2018} and it has been used to study in detail the dynamics of the KH instability during its nonlinear and turbulent stages, focusing on the kinetic effects at ion scales \citep{settino2020}.
In the present paper, we use the results described in \cite{settino2020} for the exact kinetic equilibrium to select quantities able to identify KH vortices in \textit{in situ} data.

The KH instability has numerically been studied by means of the Hybrid Vlasov-Maxwell (HVM) code \citep{valentini2007} in a 2D-3V phase space configuration (two dimension in physical space and three dimensions in velocity space), with a uniform magnetic field, $B_0$, perpendicular to the velocity shear. The Vlasov equation is integrated for the ions, while electrons are treated as a massless fluid, whose response is taken into account through a generalized Ohm's law for the electric field. The quasi neutrality condition is adopted ($n_i=n_e$) and the electron pressure is considered as a further independent quantity \citep[see][for details]{Malara2018}. The system is perturbed with a broadband spectrum of velocity fluctuations, generated in form of random noise. Finally, the ion plasma beta is $\beta_i = 2 v_{th}/v_A=2$, being $v_{th}$ and $v_A$ the ion thermal and the Alfv\'en speed, respectively; while the electron to ion temperature ratio is $T_e/T_i =1$.
The following normalization for the HVM set of equations has been used: density is normalized by $n_0$ (the density far from the shears); time by the inverse proton cyclotron frequency $\Omega_{cp}=eB_0/m_i$ (where $e$ is the electron charge and $m_i$ is the ion mass); velocity by the Alfv\'en speed $v_A=B_0/\sqrt{4\pi n_0 m_i}$; lengths by the ion skin depth, $d_i=v_A/\Omega_{cp}$; magnetic field by $B_0$, the electric field by $v_A B_0/c$ (being c the speed of light) and pressure by $ n_0 m_i v_A^2$

The unperturbed configuration is characterized by a velocity shear along the $y$-direction and varying along $x$, imposed as a hyperbolic tangent profile which has been duplicated to satisfy periodic boundary conditions in the physical space. The rotational motion, induced by the velocity shear, leads to the generation of a centrifugal force and, as a consequence, the formation of vortices along the two shear layers. During the nonlinear stage the vortices start merging, the ones at each of the shear layer and then the two shear layers interact. 
Moreover, an enhancement of the space averaged total current density occurs during the evolution of the instability, due to both the mixing of vortices at large scales and the nonlinear coupling of modes at short wavelengths.

The centrifugal force becomes stronger toward the center of the vortex, so that a pressure gradient is generated to balance it. Therefore, a local ion pressure minimum inside the vortex is recovered, as shown in the contour plot of $P_i$ in panel (a) of Figure~\ref{fig:contour}. Moreover, the strong swirling motion enhances the magnitude of the ion vorticity, namely ${|\bf \omega}_i| = |\nabla \times {\bf u}_i|$, which peaks at the edges of the vortices, but, inside them, still reaches values higher than the background vorticity
(see panel (b)).
The spatial variation of the local ion plasma beta (not shown) indicates that the ion pressure is everywhere higher than the magnetic pressure ($\beta_i > 1$) and, in particular, it peaks within the KH vortices. Therefore, the magnetic field is carried by the vortical flows and the field lines are twisted and rolled-up according to the whirling motion of the plasma. Moreover, the magnetic field lines are highly distorted in correspondence of these structures and the generation of strong 
turbulent activity is also found. 
Indeed, in panel (c) of Figure~\ref{fig:contour}, an enhancement in the magnitude of the total current density, namely $|{\bf j}|$, 
is observed at the edges of the KH vortices, while a minimum is found at the center.

During the nonlinear phase of the KH instability, the nonlinear coupling of the modes generates an energy cascade towards small scales and kinetic effects come into place, through complicated non-Maxwellian deformations. Indeed, the ion distribution functions are found to be far from thermodynamical equilibrium. Deviations from the Maxwellian shape have been quantified via the non-Maxwellian parameter, defined for a fixed time as \citep{greco2012}
\begin{equation}
\epsilon_{M} = \frac{1}{n_i}\sqrt{\int \left[f_i-g_{M}\right]^2 d^3v}
    \label{eq:eps}
\end{equation}
where $f_i$ is the actual distribution function and $g_{M}$ is the associated Maxwellian built with the moments of $f_i$. 
The contour plot of $\epsilon_{M}$, in panel (d) of Figure~\ref{fig:contour}, shows an opposite behavior with respect to $|\bf j|$. Indeed, while the total current density provides information about the boundaries of the vortices, the non-Maxwellianity peaks inside the structure, allowing the identification of the vortex core. 
Departures from the thermodynamic equilibrium are also observed in the iso-contours of both the ion temperature anisotropy ($T_\perp/T_\parallel$, where $T_\parallel$ and $T_\perp$ are the temperatures in the direction parallel and perpendicular to the local magnetic field, respectively) and the ion agyrotropy ($\sqrt{Q}$), shown in panels (e) and (f) of Figure~\ref{fig:contour}, respectively. The agyrotropy is defined as \citep{swisdak2016}

\begin{equation}
    Q=\frac{P_{xy}^2+P_{xz}^2+P_{yz}^2}{P_\perp^2+2P_\perp P_\parallel};
    \label{eq:gyrotropy}
\end{equation}
where $P_{ij}$ are the components of the ion pressure tensor in the reference frame in which one of the axes is along the local magnetic field and the two perpendicular pressures are equal. 
$\sqrt{Q}$ ranges from $0$ (fully gyrotropic configuration) to $1$ (maximum agyrotropy). We found that $T_\perp/T_\parallel$ has a pattern similar to $\epsilon_{M}$, while $\sqrt{Q}$ exhibits a behavior similar to those visible in the contour plots of $|{\bf j}|$. 
Indeed, the strongest anisotropies are observed in the center of each vortex (same as $\epsilon_{M}$), while $\sqrt{Q}$ peaks at the edges of the vortices (similar to $|{\bf j}|$).

\section{KH vortices in MMS data} \label{sec:MMS}

\begin{figure*}[ht]
    \centering
    \includegraphics[width=14 cm]{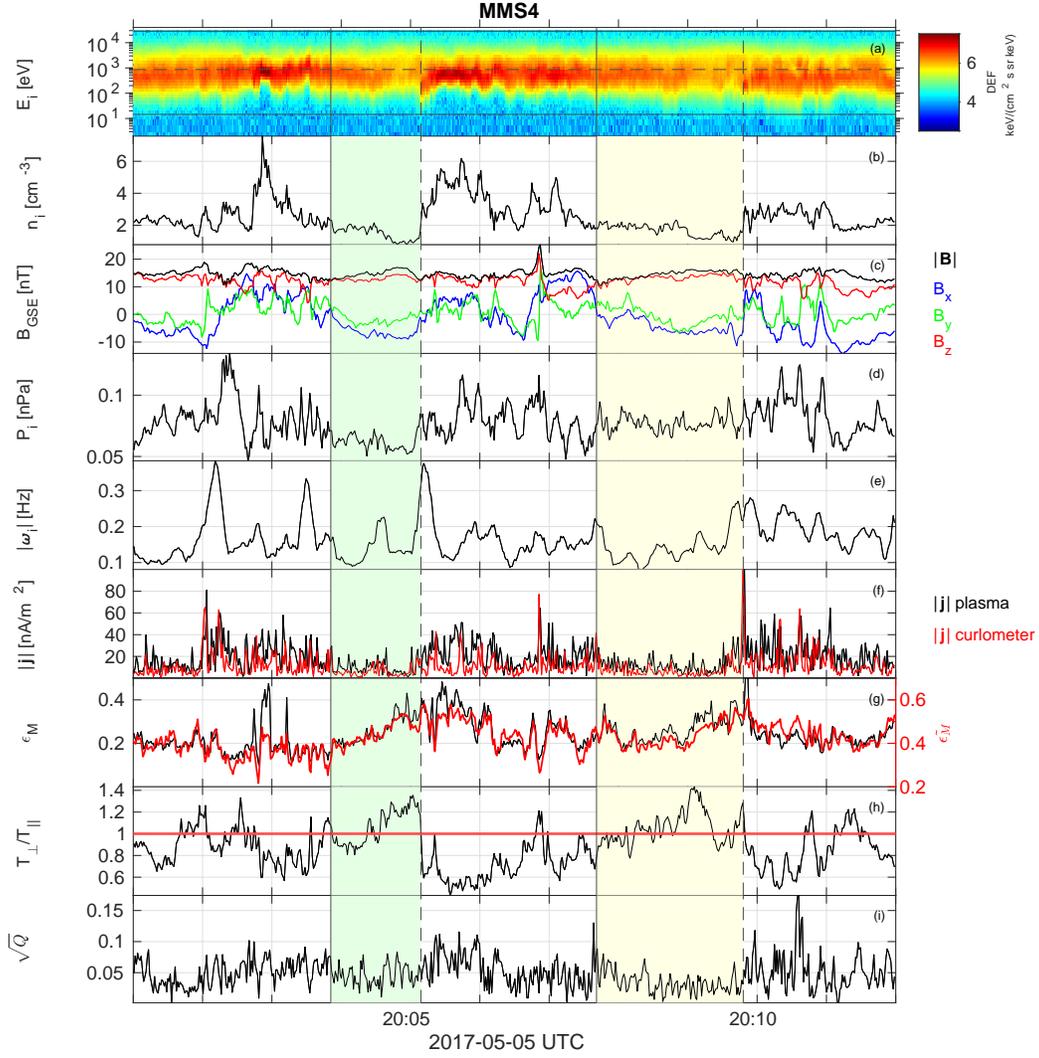}
    \caption{MMS data in the interval 20:01--20:12~UTC on May 5 2017. Measurements are averaged on $\sim$ 1s and all coordinates are in GSE. From top to bottom: (a) ion energy spectrogram; (b) ion density; (c) magnetic field; (d) ion pressure; (e) magnitude of the ion vorticity;(f) magnitude of the total current density evaluated from particle moments (black) and with the Curlometer technique (red); (g) dimensionless ion non-Maxwellianity defined as in Equation~\ref{eq:eps} (black), and Equation~\ref{eq:eps_bim} (red); (h) ion temperature anisotropy, where the horizontal red line indicates isotropic distribution function; and (i) ion agyrotropy. Colored shaded areas highlight two KH vortices and the vertical solid (dashed) lines indicate the leading (trailing) edge of the vortex.}
    \label{fig:mms4}
\end{figure*}

On the May 5 2017, between 19:30:00 and 21:00:00 UT, MMS was located at the dawn flank of the Earth during a period of mostly northward interplanetary magnetic field. In particular, MMS was at $[-14,-18,3]$ in Geocentric Solar Ecliptic (GSE) coordinates and the spacecraft separation was $\sim 156$~km. Considering that the ion inertial length, $d_i$, is $\sim 100$~km in the magnetosheath side and $\sim 230$~km in the magnetospheric side, the spacecraft separation allows detailed studies of kinetic scales. 
During this interval, MMS observed several fluctuations with a period between $2.5$ and $6$~min in different quantities, namely density, magnetic field and both ion and electron temperature, suggesting a crossing of KH vortices.
A detailed study of this event has been conducted by \cite{hwang2020} and \cite{kieokaew2020} that, indeed, interpreted such observations as the detection of KH vortices. 

In the present analysis, we focus on the time interval 20:01:00--20:12:00~UTC to investigate the behavior of the significant quantities for the  identification of KH vortices highlighted by the HVM simulation.
We use magnetic field data from fluxgate magnetometer \citep[][]{russell2016} sampled at $128$Hz, and the ion data from Fast Plasma Investigation instrument \citep{pollock2016} with a resolution of $150~ms$. An overview of the interval is shown in Figure~\ref{fig:mms4}, where the crossing of the two vortices by MMS4 has been marked by the green and yellow shaded areas. Moreover,
leading and trailing edges are indicated by vertical solid and dashed lines, respectively, corresponding to the times provided by \cite{hwang2020}.

The energy spectrogram of ions (panel a) displays the crossing of a mixing layer by the spacecraft.
Inside the vortices (shaded areas) a less mixed region of more magnetospheric-like ions can be recognized. This feature is in agreement with the crossing of a leading edge of the vortex because of the centrifugal force which tends to confine the low density and high energy plasma (magnetosphere-like) close to the vortex core \citep{hasegawa2004a,hwang2012,hwang2020}.  
The same behavior can be recognized in the ion density, $n_i$, shown in panel (b). Indeed, two clear minima are observed in correspondence with each vortex and close to the trailing edge (vertical dashed lines). It is worth pointing out that the presence of such low density regions indicates a low number of particle counts inside the vortices and, thus, a possible increase of uncertainty in the particle measurements. Therefore, in order to improve the counting statistics and to have reliable observations, we averaged all the quantities shown in Figure \ref{fig:mms4} over seven time steps (roughly $1$~s), which corresponds in terms of spatial scale to $\sim d_i$.

Panel (c) shows the magnetic field in the GSE coordinate system. It can be easily seen that $\bf B$ is directed mostly northward along the whole interval (black and red lines), while an inversion in the sign of $B_y$ and $B_x$ are  found, in agreement with the crossing of vortex boundaries.
Panel (d) shows the ion pressure, $P_i=k_B n_i T_i$, where $k_B$ is the Boltzmann constant and $T_i$ is the ion scalar temperature.
The pressure inside the first vortex (green shaded area) is lower than the surrounding regions, while for the second vortex (yellow shaded area) we do not see this decrease in pressure. This different feature can be connected to the presence, before the second vortex, of a Flux Transfer Event (FTE) as reported by \citet[][]{kieokaew2020}, and/or to a high value of the background pressure. Indeed, the FTE determines an increase of the magnetic pressure, playing the main role in balancing the centrifugal force.

In panels (e) and (f) we show the ion vorticity, ${\bf \omega}_i$, and the total current density, ${\bf j}$, respectively, estimated through multi-spacecraft techniques. In particular in panel (f) we compare the magnitude of the current density, by using both the single-spacecraft plasma measurements ${\bf j}=en_i({\bf V}_i-{\bf V}_e)$ (black line), where the quasi-neutrality assumption has been used, and the Curlometer technique (red line) \citep{dunlop1988,dunlop2002}. Since the behavior of these two quantities is similar, we are confident that the multi-spacecraft technique works well. Thus, we estimate the ion vorticity in the same manner as in \cite{perri2020}. We find that $|{\bf \omega}_i|$ peaks at the boundaries of the vortices, although strong spikes are also observed inside. 
It is worth pointing out that strong spikes are observed along the whole interval suggesting that the level of the background vorticity (due to the presence of the velocity shear) is comparable to the vorticity enhancement connected to the rotational motion. Thus, it is difficult to distinguish vortex boundaries by using only this quantity.  
On the contrary, a clearer behavior is found for the magnitude of the total current density in panel (f). We find strong spikes at the edges of both the vortices, while low values are generally observed inside them.

Figure~\ref{fig:mms4}g shows the departure from a Maxwellian distribution, ${\epsilon}_{M}$. We use the same definition as in Equation~\ref{eq:eps}, multiplied by $v_A^{3/2}$ (being $v_A$ the Alfv\'en speed in the magnetosheath) in order to have a dimensionless quantity. Moreover, for completeness, we use a second definition for the non-Maxwellianity, recently introduced in the framework of MMS observations (Graham et al., 2021, \textit{in prep.})
\begin{equation}
    \tilde{\epsilon}_{M} = \frac{1}{2n_i} \int \left|f_i-g_{M}\right| d^3v,
    \label{eq:eps_bim}
\end{equation}
a dimensionless quantity that ranges from 0 (Maxwellian distribution) to 1 (highest deviation from a Maxwellian). 
To reduce the artificial increase of non-Maxwellianity due to the noise associated with the lowest energy channels, we have integrated, both $\epsilon_M$ and $\tilde{\epsilon_M}$, from $15$~eV to $30$~keV (see horizontal lines in panel (a) of Figure~\ref{fig:mms4}).
Furthermore, to reduce the impact of the low counts statistics inside the vortices, we have estimated both $\tilde{\epsilon}_{M}$ and $\epsilon_{M}$ by averaging the ion distribution function on $\sim 1$~s. Finally, we have also verified that averaging the ion distribution function over longer times does not significantly change the results.
Although non-Maxwellian distributions are observed along the whole time interval (since the plasma is almost non collisional), we also observe significant displacements from a Maxwellian distribution, higher than the mean value, in the shaded areas, with the peaks close to the trailing edge of each vortex.
$\tilde{\epsilon}_{M}$ (red line) and $\epsilon_{M}$ (black line) display a similar trend. Indeed, the correlation between the two non-Maxwellianity parameters is very high, with a correlation coefficient of $\sim 0.8$.

\begin{figure*}[ht]
\centering
\begin{minipage}[hb]{0.45\linewidth}
   \centering
\includegraphics[width=0.94\textwidth]{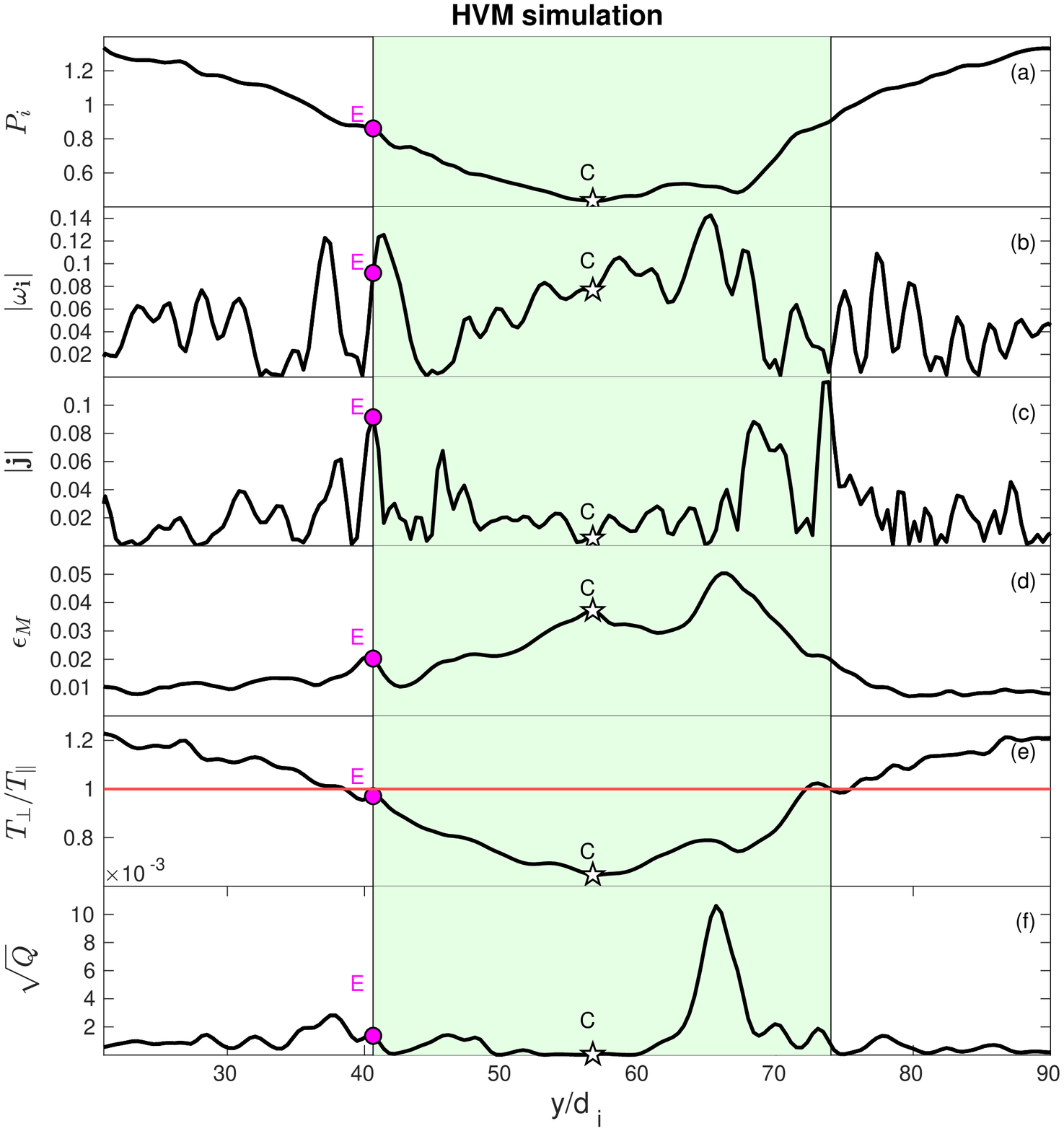}
\end{minipage}
\begin{minipage}[hb]{0.45\linewidth}
   \centering
\includegraphics[width=0.94\textwidth]{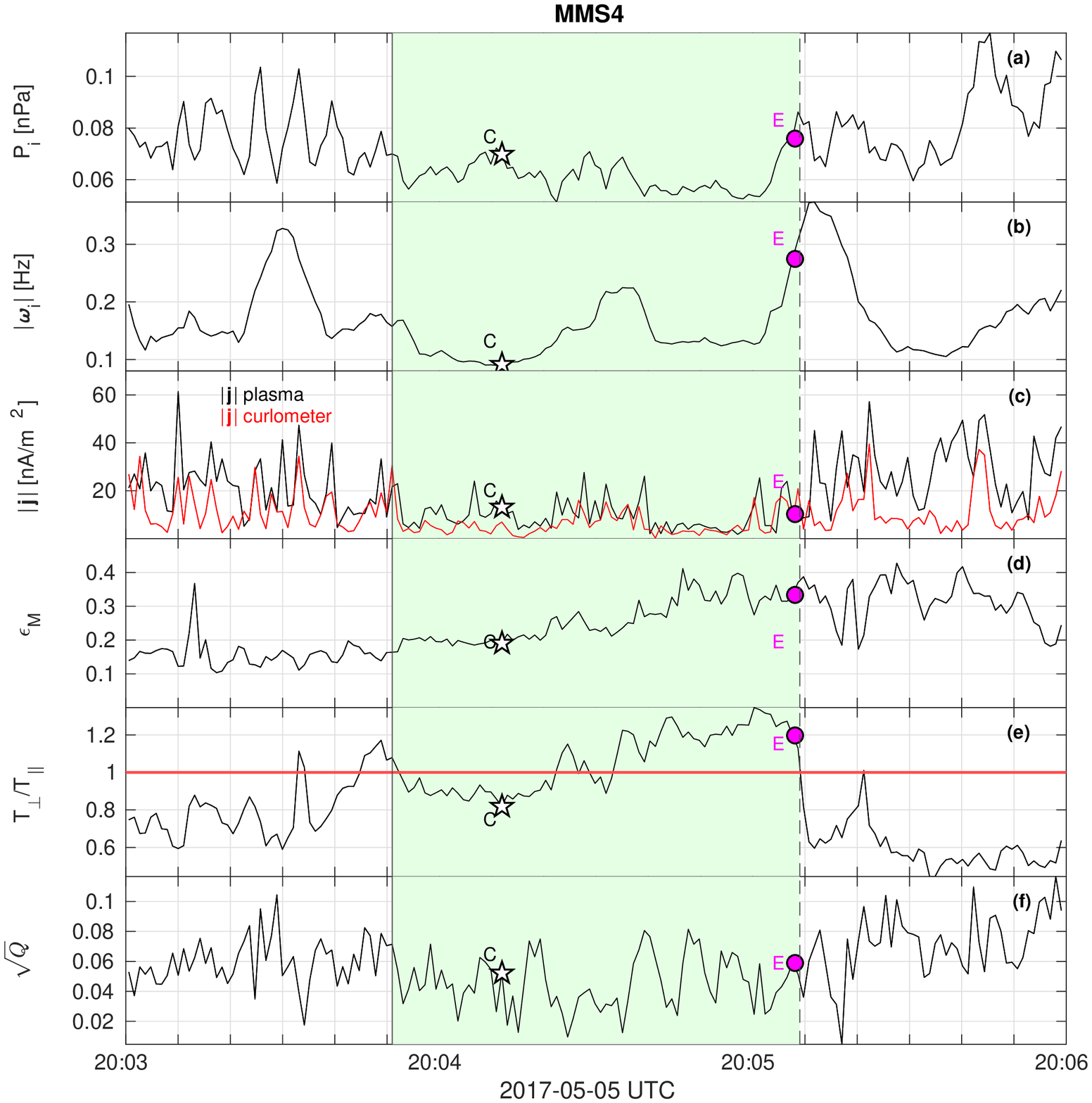}
\end{minipage}

    \caption{(Left) simulation: one-dimensional cuts of the quantities shown in Figure \ref{fig:contour}, along vertical dashed lines at $x=53.52 d_p$. (Right) MMS: zoom of the data in Figure \ref{fig:mms4}. From top to bottom: (a) ion kinetic pressure; (b) magnitude of the ion vorticity; (c) magnitude of the total current density; (d) ion non-Maxwellianity; (e) ion temperature anisotropy; and (f) ion agyrotropy. The green shaded area highlights the vortex, while the magenta circles and the white stars indicate the points selected for the investigation of the ion distribution function at the edge (E) and inside (C) the vortex. }
    \label{fig:cut}
\end{figure*}

Finally, in panels (h) and (g) of Figure~\ref{fig:mms4}, we plot the ion temperature anisotropy, $T_\perp/T_\parallel$, and the ion agyrotropy, $\sqrt{Q}$, respectively. The definition for these quantities is the same used in Section~\ref{sec:HVM}.
Anisotropic distribution functions are observed inside the vortices, while strong agyrotropy is found near the edges, particularly evident for the second vortex (yellow shade). Moreover, the change of the anisotropy direction in panel (h) defines the boundary of the vortices. Indeed, it passes from a parallel ($T_\perp/T_\parallel<1$) to a perpendicular ($T_\perp/T_\parallel>1$) anisotropy or vice versa, at the edge of the vortices (red horizontal line indicates isotropy).
This is more evident for the first vortex (green shaded area), while for the second one (yellow shaded area) this change is not observed at the leading edge, which may be connected to the presence of the FTE.
It is worth noting that, inside the vortices, both parallel and perpendicular temperature anisotropy is observed.

\section{Discussion}\label{sec:discussion}

In this section we compare the results from the hybrid KH simulation (Section \ref{sec:HVM}) with MMS observations (Section \ref{sec:MMS}). It is worth pointing out that, although a multi-spacecraft mission has been used for the present work, the quantities we suggest for the identification of the vortices can be easily estimated with high-resolved single-spacecraft measurements. Moreover, since in the simulation we focus on a vortex at a fixed time during the nonlinear phase of the instability, we assume that the vortex observed by MMS does not evolve during the spacecraft crossing. Indeed, the large scale vortices are propagating faster than the spacecraft and the Taylor hypothesis is therefore valid. 
To be more quantitative, the spacecraft speed, $V_{\text{sc}}$, is of the order of $2$~km/s, while the vortex speed can be evaluated as defined by \citet{otto2000}:
\begin{equation}
 V_{\text{vort}} = \frac{V_i}{2}\frac{n_{\text{MSH}}-n_{\text{MSP}}}{n_{\text{MSH}}+n_{\text{MSP}}},
 \label{eq:v_vort}
\end{equation}
where $V_i$ is the ion bulk velocity averaged in the time interval analyzed and $n_{\text{MSH(MSP)}}$ is the ion density averaged in the magnetosheath  (magnetospheric) side. In our case $V_{\text{vort}}=120$~km/s ($\gg V_{\text{sc}}$).
Since the spacecraft can be considered at rest, we can assume that vortices are observed in the same evolutionary phase.
Moreover, the evolution of the KH vortices generally occurs while they are propagating towards the tail.

Since \textit{in situ} measurements only provide information on a one-dimensional trajectory in time/space, we have decided to choose a 1D path of the numerical quantities which crosses the center of the vortex, i.e. $x=53.52~d_i$ in Figure \ref{fig:contour} (vertical dashed lines) compared with the MMS4 observations (right panels).
Moreover, for a one-to-one comparison, we have decided to consider, for the MMS observations, only the crossing of the first vortex (green shading in Figure~\ref{fig:mms4}). The results for the simulation (left) and MMS (right) data are shown in Figure~\ref{fig:cut}, where the vortex regions are highlighted by the green shaded areas.

\begin{figure*}[ht]
\centering
  \includegraphics[width=\textwidth]{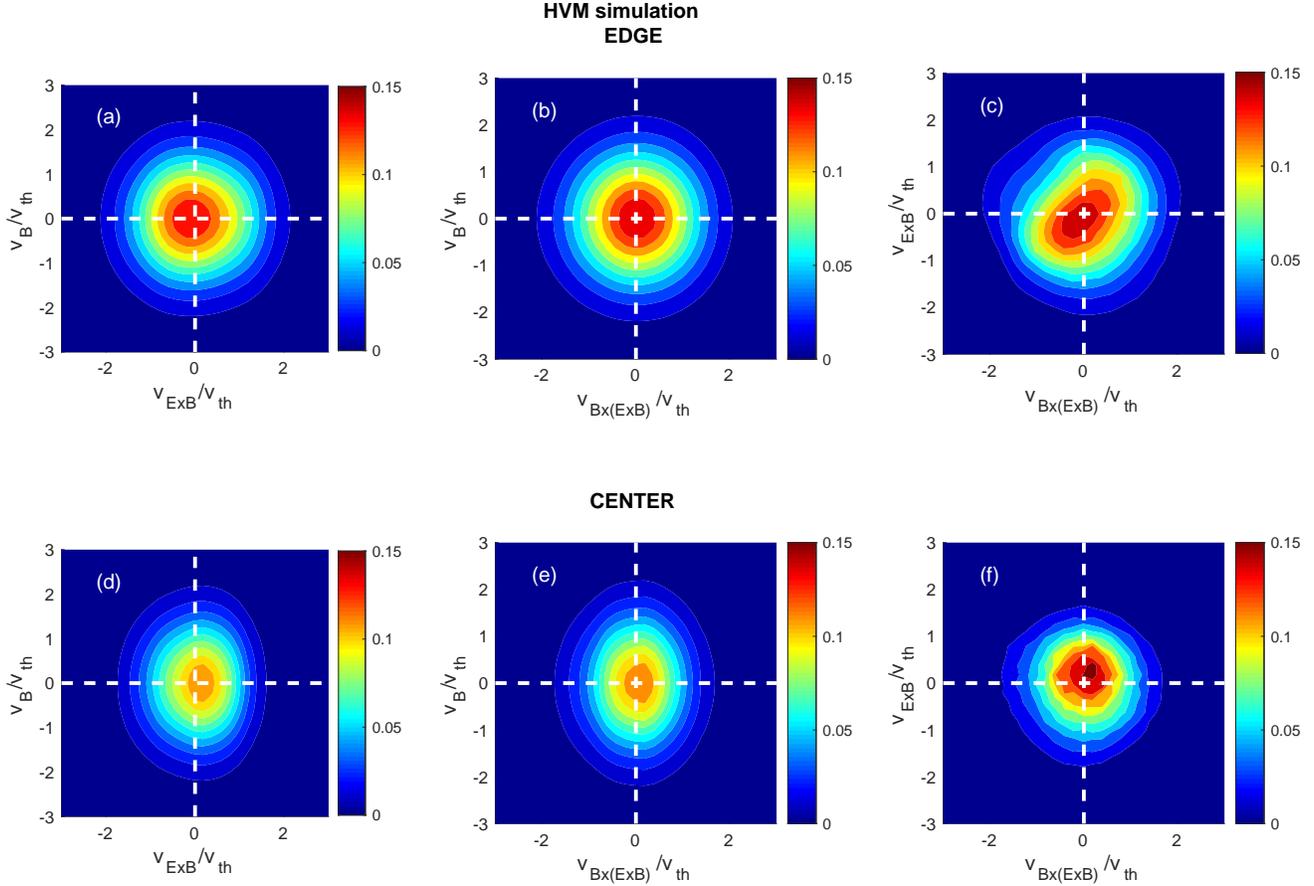}
\caption{Two-dimensional contour plots of the reduced ion distribution function at $E=(53.52,40.625)d_i$ (top row) and at $I=(53.52,56.64)d_i$ (bottom row). From left to right, the ion velocity distribution is shown in the planes $(v_{{\bf E}\times {\bf B}},v_{\bf B})$, $(v_{\bf{B}\times(\bf{E}\times \bf{B})},v_{\bf B})$, and $(v_{\bf{B}\times(\bf{E}\times \bf{B})},v_{\bf{E}\times \bf{B}})$, respectively. }
    \label{fig:vdf}
\end{figure*}

\begin{figure*}[ht]
\centering
\begin{minipage}[ht]{\linewidth}
   \centering
\includegraphics[width=\linewidth]{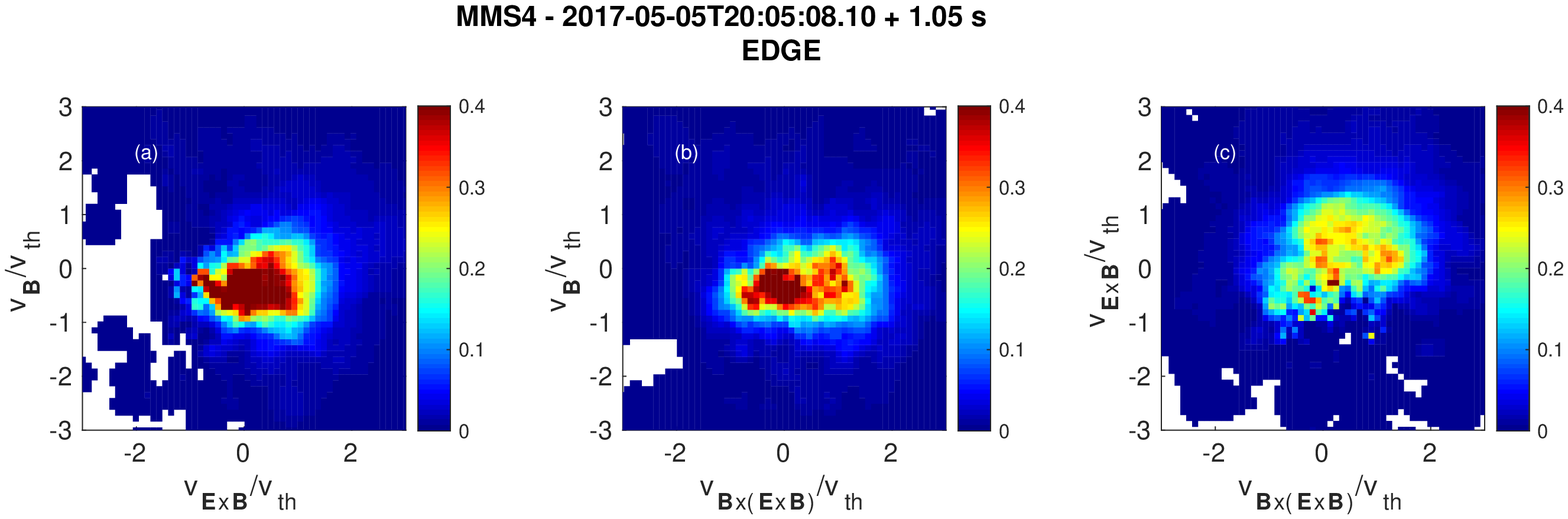}
\end{minipage}
\begin{minipage}[ht]{\linewidth}
   \centering
\includegraphics[width=\linewidth]{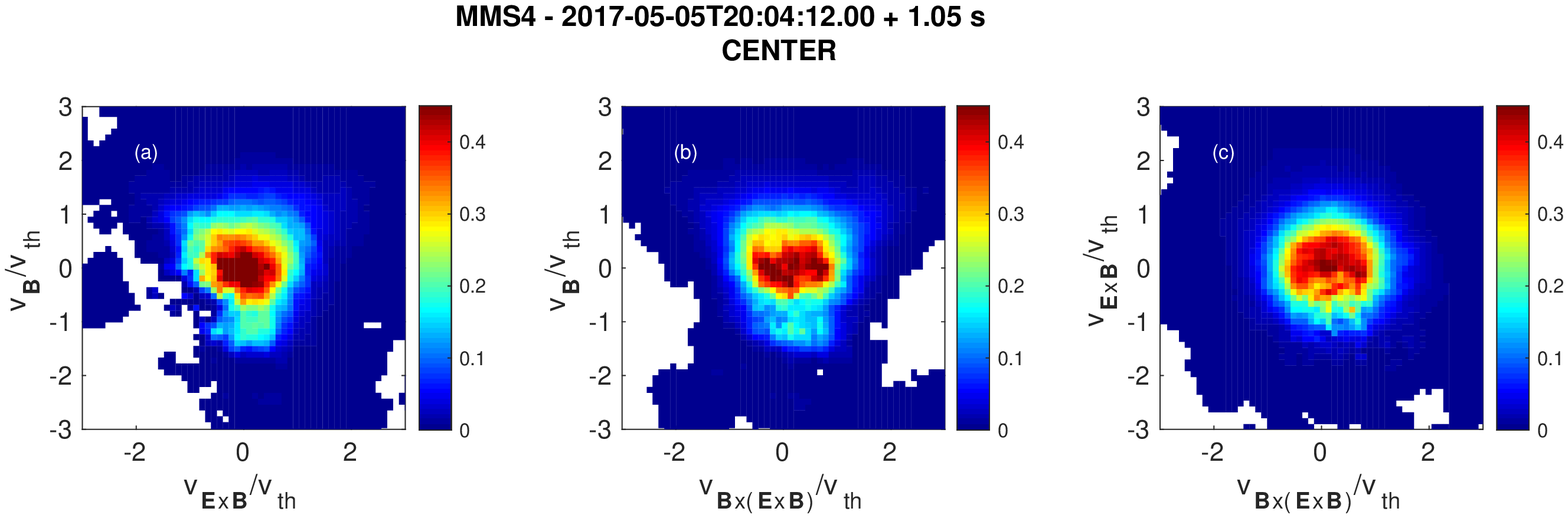}
\end{minipage}
\caption{Two-dimensional reduced ion velocity distributions at the edge (top row) and inside (bottom row) the vortex highlighted in Figure~\ref{fig:mms4} by the green shadow. The same reference frame used for the simulation results in Figure~\ref{fig:vdf} has been chosen.}
    \label{fig:vdfMMS}
\end{figure*}

Both HVM simulation and \textit{in situ} data show a local minimum for the ion kinetic pressure, $P_i$, within the vortex (see panels a), in agreement with the physical mechanism that produces this structure. 
Indeed, the rotational force leads to the generation of a pressure gradient which balances the force itself. 
However, while the pressure minimum is clearly visible in the HVM simulation, much more fluctuations are observed in the MMS data. In addition, as discussed in the previous section, no pressure minimum is observed in the second vortex (yellow shade in Figure \ref{fig:mms4}), suggesting that $P_i$ is not a good parameter to identify KH vortices in space observations. 
A similar argument can be used for the vorticity shown in panels (b). Indeed, for the HVM simulation, evident peaks at both the edges and inside the vortex are observed. In the MMS case, instead, strong spikes are not limited to the vortex region but are observed all along the interval.

Panels (c) show the magnitude of ${\bf j}$. We found that the edges of the vortex are not a thin structure, but instead display a multiple filamentary structure (already observed in Figure~\ref{fig:contour}{c}).
High values of  $|{\bf j}|$ are observed at the edges of the vortex and even if few small spikes can be recognized inside the shaded region, they still remain lower than the outside region. Moreover, really low values are reached close to the trailing edge of the vortex (vertical dashed line) where the non-Maxwellianity is high (panel d on the right). Indeed, $\epsilon_M$ displays an opposite behavior respect to $|{\bf j}|$, since it is enhanced inside the vortex, while decreases in correspondence of the boundaries. 

At the edge of the vortex, the transition from the outside region to the vortex structure is well represented by a change in the direction of the ion temperature anisotropy (see panels e). However, while in the HVM case we pass from $T_\perp > T_\parallel$ to $T_\perp < T_\parallel$ (crossing the red horizontal line, which identifies isotropy), in the MMS crossing the inverse is observed. 
Moreover, while numerical results show $T_\parallel > T_\perp$ in the center of the vortex, MMS observe either parallel and perpendicular temperature anisotropies inside the green shade. 
In Figure~\ref{fig:cut}f, we show the ion agyrotropy which is characterized by a similar behavior with respect to $|{\bf j}|$. Indeed, the highest values are observed at the edges, while a minimum is found inside the vortex region.

The comparison between hybrid simulation and MMS measurements suggests that the combined use of the non-Maxwellianity parameter and of the magnitude of total current density can preliminary identify the KH vortex region and provides information about the edges and the center of the vortices. Moreover, the peaks in the agyrotropy and the change of direction in the temperature anisotropy can be used for the identification of the outer boundaries of the vortex.
At this stage, it is interesting to consider the effects of the KH instability on the ion velocity distribution function (VDF), focusing on two distinct locations in physical space, namely the outer edge of the vortex, where $|\bf j|$ has a peak while $\epsilon_M$ is low, and the center of the vortex, in correspondence of an enhancement of $\epsilon_M$ and the minimum of the total current density. These two points, $E$ and $C$, have also been indicated in both Figures~\ref{fig:contour} and \ref{fig:cut}  as a magenta circle and a white star, respectively.

In Figure~\ref{fig:vdf}, we show the two-dimensional contour plots of the numerical ion VDF at $E$ (top) and $C$ (bottom), integrated along the out-plane direction, where the velocity grid has been normalized to the ion thermal speed. From left to right, the VDF is shown in the planes $(v_{\bf E \times B}, v_{\bf B})$, $(v_{\bf B\times(B\times E)},v_{\bf B})$, and $(v_{\bf B\times(E\times B)},v_{\bf E\times B})$, respectively.
The ion VDF appears to be strongly distorted both at the edge and center of the vortex, but different features can be recognized. In particular, at the edge of the vortex, a strong non-gyrotropic VDF is observed (panel c), with a significant elongation in the oblique direction. 
Moreover, while in panel (b) the VDF is mostly isotropic, in panel (a) a small beam around $v_{th}$ can be observed in the direction perpendicular to the magnetic field. 
The presence of these accelerated particles could be generated by the sharp changes in the magnetic field that characterize the vortex boundaries. The field lines are highly distorted by the rotational motion of the plasma and, as a consequence, strong currents are generated.
On the other hand, at the center of the vortex, the VDF is almost gyrotropic (see panel f) and is significantly elongated in the direction parallel to the magnetic field (see both panels d and e).

In Figure \ref{fig:vdfMMS} we show the reduced ion VDF observed by MMS4 at both the edge of (top) and inside (bottom) the vortex. The time at which the two VDFs have been picked is also indicated in the right panels of Figure~\ref{fig:cut} with a magenta circle (edge of the vortex) and a white star (center of the vortex).  
We averaged the VDFs over seven time steps to improve the counting statistics. In the chosen averaging time interval, all the VDFs have the same behavior, since they lie in regions with the same characteristics, i.e. high $|{\bf j}|$ and low $\epsilon_{M}$ at the edge of the vortex (and the opposite in the center).

 At the edge of the vortex (top) some common features with the numerical simulation can be observed (see Figure \ref{fig:vdf}). The ion distribution function is strongly agyrotropic, with an elongation in the oblique direction (panel c).  It is worth pointing out that high non-gyrotropic signatures have also been observed  in the electron distribution function in correspondence of the boundaries of KH vortices in a fully kinetic KH simulation, while a mostly gyrotropic distribution has also been found inside the vortex \citep{nakamura2020}. The presence of the same trend found in KH vortices at electron scales and in different simulations suggests that such feature is independent of both scalelength and initial conditions.
 Moreover, in panel (a) a clear beam can be observed in the plane perpendicular to the magnetic field and along the positive direction. Here, we argue that in our simulation the beam at the thermal speed seems less evident because of the high value of the ion plasma beta ($\beta_i=2$) which means a more spread for the core of the ion distribution, compared with the average $\beta_i=0.8$ observed by MMS4.
 
 Besides these analogies, we underline that at the edges of the vortex, MMS detected a strong perpendicular temperature anisotropy in the direction $v_{\bf{B}\times(\bf{E}\times \bf{B})}$ (panel b), in contrast with the mostly isotropic distribution function observed in the numerical simulation. This is probably due to the transition from magnetospheric-like to magnetosheath-like particle populations in correspondence of the edges of the vortices, not described in our simulation.
 
 Finally, at the center of the vortex (bottom), we find a good agreement with the HVM simulation. We observed an almost gyrotropic ion distribution and a significant elongation is found in the direction parallel to the magnetic field. However, unlike what is found in the simulation, where a parallel temperature anisotropy is present in the whole vortex region, inside the vortex crossed by MMS also distributions with significant elongation in the direction perpendicular to the local magnetic field can be observed.

\section{Conclusions}\label{sec:summary}

In this paper, we have compared \textit{in situ} data of a KH event and results from a numerical simulation, suggesting new quantities to help the detection of KH vortices in single-spacecraft measurements. A very recent hybrid Vlasov simulation of KH instability~\citep{settino2020} has shown some interesting features which arise at kinetic scales. Motivated by these clear small-scale signatures, we have decided to test for their presence in a KH event observed by MMS for which the trailing and leading edges of the vortices had already been recognized \citep{hwang2020}. We point out that our simulation has been performed in different initial conditions respect to the MMS event, but nonetheless the results are in reasonable agreement, being, the identified patterns, an intrinsic characteristic of the vortices.

We found, in both synthetic and \textit{in situ} data, a clear enhancement of the magnitude of the total current density at the edges of the vortex, followed by a minimum near the center of the vortex itself. A opposite behavior is observed for the ion non-Maxwellianity, which peaks inside the vortex, denoting a high distortion of the distribution function, and is low at the edges.
Moreover, in correspondence of these strong non-thermal features, also temperature anisotropy and agyrotropy are observed. In particular, we found that the ion agyrotropy behaves similarly to the magnitude of the total current density, since it reaches a local minimum inside the vortex, while increases at the boundaries. A change in the direction of the ion temperature anisotropy is also observed at the edges of the vortex.

The main techniques to detect KH vortices in plasmas are based on multi-spacecraft approaches \citep{hasegawa2004a,cai2018}, while single spacecraft techniques have strong limitations \citep{jeong1995,kida1998,plaschke2014}.
For instance, techniques based on local pressure minima and enhancement in the vorticity magnitude fail when the background fluctuations are comparable to the value inside the vortex or when the dynamics of the KH instability is affected by other ongoing phenomena. The inefficiency of such quantities, namely the kinetic pressure and the vorticity magnitude, is also recovered in our analysis, since a well defined minimum of the ion kinetic pressure is observed inside the KH vortex of the hybrid simulation (Figure~\ref{fig:cut}a), in agreement with the $3$D fully kinetic PIC simulation in southward interplanetary magnetic field condition \citep{nakamura2020}, but not in MMS data (Figure~\ref{fig:mms4}a).
Indeed, a flux transfer event has been observed in the MMS interval we considered~ \citep{hwang2020,kieokaew2020}, suggesting that the absence of a pressure minimum is connected to the increase of the magnetic pressure.

Our analysis suggests that the investigation of the total current density, non-Maxwellianity and both temperature anisotropies and agyrotropy enables the identification of KH vortices in space measurements, requiring only a single satellite and a good resolution for particle instruments. 
Therefore, we suggested that these quantities could be also used in the framework of the new ESA's Solar Orbiter mission \citep{muller2020}, launched in February 2020, to investigate the presence of KH vortex structures at the interaction region between fast and slow solar wind, providing a significant insight on the instability and more generally on kinetic effects in the near-Sun solar wind.    

\section*{Acknowledgments} Numerical simulations have been run on Marconi supercomputer at CINECA (Italy) within the ISCRA projects: IsC68\_TURB-KHI and IsB19\_6DVLAIDA. This work has received funding from the European Unions Horizon 2020 research and innovation programme under grant agreement no. 776262 (AIDA, \url{www.aida-space.eu}).
The data used in this paper are freely available from the MMS data center
(\url{https://lasp.colorado.edu/mms/sdc/public/}).

\bibliography{bibliography}

\end{document}